\documentclass[aps,groupedaddress,showpacs,floatfix]{revtex4}
\usepackage{graphics}
\usepackage{epsfig}
\begin{document}
\title{Development of an interatomic potential for phosphorus impurities 
in $\alpha$-iron}
\author{G.J.Ackland$^1$, 
M.I.Mendelev$^2$,  D.J.Srolovitz$^2$, S.Han$^3$ and A.V.Barashev$^4$  }
\affiliation{
$^1$ School of Physics,The University of Edinburgh, Mayfield Road, Edinburgh, EH9 3JZ, UK,
$^2$ Princeton Materials Institute
Princeton University, Princeton, NJ 08544 USA
$^3$ Department of Physics,
Ewha Womans University, Seoul 120-750 Korea.
$^4$Department of Engineering,
The University of Liverpool,
Liverpool, L69 3GH,
UK}

\begin{abstract}
We present the derivation of an interatomic potential for the 
iron phosphorus system based primarily on {\it ab initio} data.  
Transferrability in this system is extremely problematic, and the
potential is intended specifically to address the problem of radiation 
damage and point defects in iron containing low concentrations of 
phosphorus atoms.  Some preliminary molecular dynamics  
calculations show that P strongly affects point defect migration.

\end{abstract}
\date{\today}
\maketitle

\section{Introduction}

Phosphorus is one of the major causes of embrittlement of
nuclear reactor pressure vessel (RPV) steels\cite{nishiyama}.  
A huge enhancement of the concentration of
phosphorus (P) atoms at grain boundaries is
observed in samples of RPV steels taken from
nuclear reactors\cite{busw}.  This leads to a decrease in the 
grain boundary cohesion
and consequently to a shift of the ductile-to-brittle transition
temperature.

A theoretical
understanding of the problem requires a full understanding of the
interaction of millions of these two atoms at 
the atomic level.
First principles calculations provide the most reliable way to describe 
interactions, but they are impractical for large scale molecular dynamics.  
Hence there is a need for accurate descriptions of the energy based 
on interatomic potentials which do not treat the electrons explicitly. 

A crucial aspect of atomic-level modelling 
is proper relaxation of atoms around the
defect, for which elegant methods were developed by Michael
Norgett\cite{norgett}.  With state-of-the-art {\it ab initio} methods
it is now possible to treat a few hundred movable atoms: just as with
empirical potentials in the seventies this brings problems of finite size
effects to the fore and many of the concepts pioneered by Michael Norgett and
encapsulated in his DEVIL\cite{devil} (Defect EValuation In Lattices)
code are being revisited today, and indeed many of the applications
such as dislocation core structure\cite{nps} and twin
boundaries\cite{bristowe} are still debated.

Here we present the derivation of an interatomic potential for the Fe-P which can be evaluated as quickly as a short
ranged pairwise potential.  For molecular dynamics purposes in
studying reactor steels, the interesting region is that of small
concentrations ($\sim 10^{-3}$) of P in Fe, 
in particular, the behaviour of point defects in lattices.
Traditionally, data for fitting potentials came from experiment,
typically bulk properties, but recently {\it ab initio} total energy
calculation enables us to add more specific atomistic level
configurations to the fitting dataset.  
Here we incorporate both ab initio and experimental data to
parameterise a potential for the dilute P-Fe system.

We choose to write the potential in the form:

\begin{equation}
U = \sum_i   F_i[\sum_j \phi(r_{ij})] + \sum_{ij} V(r_{ij})
\end{equation}

For $F_i[x]=\sqrt{x}$ this is the second-moment 
tight-binding form of Cyrot\cite{cyrot} and Finnis and Sinclair\cite{FS}.
This is transferable between environments where the local band structure
is primarily changed by scaling\cite{valid}.  
We will need to fit seven
functions for the two-component system: three pair potentials $V_{FeFe}(r)$,
$V_{FeP}(r)$ and
$V_{PP}(r)$,
three pairwise functions 
$\phi_{FeFe}(r)$,
$\phi_{FeP}(r)$ and
$\phi_{PP}(r)$,
and two embedding energy functions $F_{Fe}$ and  $F_{P}$.

 A previous parameterization in this form for general properties of
iron\cite{Fe} gave good qualitative results but was not tailored
specifically to defect properties.  Recent ab initio work\cite{HanVMo,mimphilmag}
shows that at high densities the resistance to compression arises
from a many-body effect (electronic kinetic energy) rather than a
pairwise repulsion.  Fitting compression data using the short range
part of $V(r_{ij})$ typically leads to an overestimate of
interstitial formation energies and volumes\cite{ATVF}.  These quantities have
recently come into the realm of what can be calculated by ab initio
means, and can be used to fit short range repulsion in regions where
the density is close to its bulk value. We have shown that including
different things in the fit leads to very different potentials: to 
describe point defect interactions, one must fit point defect properties. and
so we use an iron potential optimised for point defects as the basis of the
present work\cite{mimphilmag}.
This incorporates the
kinetic energy effect by writing $F(x)=-\sqrt{x}+a_2x^2+a_4x^4$, with 
increasing $x$ taking us smoothly between hopping and free-electron 
dominated regimes.  Hence isotropic compression is dealt with by the 
many-body term (high $x$), while short atom-atom distances can be 
addressed with the pair potential $V(r_{ij})$.  This modest change to the 
second-moment formalism gives greater fitting flexibility and 
no additional computing cost, since $F$ is implemented as 
a look up table.

Pure P is 
covalently bonded and cannot be described by this type of potential. 
We therefore do not 
attempt to fit properties of pure phosphorus, or phosphorus-rich compounds, 
concentrating instead on point defects in $\alpha-$iron, their interactions
with phosphorus atoms and fictitious iron-rich compounds.  
To parameterise the potential we use data generated from {\it ab
initio} plane wave pseudopotential calculations using density
functional theory\cite{KS}, ultrasoft
pseudopotentials\cite{vanderbilt}, the spin-dependent generalised
gradient approximation for exchange and correlation\cite{gga} and
periodic boundary conditions.  Calculations on pure iron\cite{domain}
suggest that this is a reliable approach, it has been deployed widely
and the calculations can be routinely done using standardised
software\cite{packages}.  

There is an additional caveat for ferromagnetic materials: The
magnetism (and hence Fermi level) is affected by defects and this
leads to a much slower convergence of the energy with system size (and
k-point sampling) than is typically observed for non-magnetic
elements\cite{SeungWuV}.  This may be due to the fact that the Fermi
energy moves relative to the spin-bands in the supercell calculation,
whereas for a truly infinite crystal the bands are fixed.  Despite
this slow convergence of defect energy, the energy differences between
e.g. different interstitial configurations converge much more
rapidly\cite{domain}.  Thus we are justified in fitting energy
differences from calculations using relatively small unit cells, while
obtaining the absolute formation energy from larger calculations.

Ferromagnetism also means that the band structure changes dramatically at 
the fcc-bcc phase transition iron.  For this reason\cite{valid} we do 
not expect the potential to describe paramagnetic iron correctly: 
high-T fcc-bcc transitions have been observed with other 
iron potentials but
the driving force is vibrational (entropic) not magnetic \cite{caro}.

\section{Methods}

The fitting strategy assumes that phosphorus interacts with point defects
via short-ranged, pairwise interactions and long-range strain fields.  Thus 
we include configurations representing point defects, single substitutional 
impurity (SSI) phosphorus atoms, and combinations thereof.  Since the major problem 
associated with phosphorus is segregation to grain boundaries, we include 
Fe with a 2D layer of P in our fitting database.  
By fitting relaxation volumes, we capture long-range strain effects. 
We also include liquid configurations to ensure that the functions are sampled 
at all separations\cite{mim}.

The ab initio calculations and configurations included in the fitting 
are given in tables\ref{tab:crystal},\ref{tab:INT},\ref{tab:v}. 
Our ab initio calculations use small unit cells
which introduces problems of images and relaxation\cite{devil} 
for considering isolated defects.  However, by fitting {\it the same} 
small cells described by the potential, we alleviate this problem.

The most crucial aspect for radiation damage is the geometry of the
defects and barriers (which governs diffusion mechanisms) and the
energy differences between them (which governs diffusion rates).
``Geometry'' here includes the symmetry of the defects and their
P-composition, but not detailed interatomic separations.  Similarly
important are the interaction strengths between phosphorus and point
defects, which determine whether pinning of defects occurs.  Of
secondary importance are the actual values of the formation energies:
this affects production rates in cascade simulations, but in a
molecular dynamics run defects are very unlikely to be generated
thermally.

\subsection{{\it Ab initio} Calculations}

Our {\it ab initio} calculations are done using standard codes\cite{packages}, implementing the pseudopotential plane wave method using density
functional theory\cite{KS}, ultrasoft
pseudopotentials\cite{vanderbilt}, the spin-dependent generalised
gradient approximation for exchange and correlation\cite{gga} and
periodic boundary conditions.  The plane wave cutoff of 300eV has been used throughout with $k$-point convergence to 0.01eV, force convergence to 0.001eV/A and strees to 0.01kB.

Previous ab initio calculations on iron-phosphorus\cite{wu,tang,zhong,sagert} 
using the similarly-reliable 
full-potential linearized augmented plane wave method have downplayed 
total energy
and concentrated on the role of increased electron density as indicating 
strengthening of the bonds at the surface relative to the grain boundary.
This qualitative picture is not specific to FeP and follows from 
the tight-binding formalism\cite{cyrot}. 
We observe similar increased electron densities
in our calculations of a single layer of
P in Fe, but do not use them in fitting.

The actual energies calculated by DFT relative to the free atom 
are known to be unreliable.  Likewise, the Finnis Sinclair formalism 
cannot be expected to be transferrable to free atoms.  Thus we do not 
at any stage fit the ab initio energies directly, rather, we fit relative 
energies between different configurations of the same number and type of 
atoms.
We take the experimental cohesive energy for $\alpha$-iron  (4.316eV/atom), 
but the cohesive energy for phosphorus is not fitted (and nor 
is the crystal structure of pure phosphorus).  All fits are to 
energy differences with phosphorus in various locations, with the single substitutional impurity taken as the reference state.

\subsubsection{Crystal structures - (table 1)}
 
Several crystal structures were examined and used in the fitting.  At
the Fe$_3$P composition we looked at fictitious L1$_2$ (fcc
equivalent), DO$_3$ and DO$_{32}$ (bcc equivalents) and a bcc-based
structure with every fourth (001) layer replaced by phosphorus.
DO$_3$ was noticably lower in energy.  

We also calculated Fe$_2$P, a complex structure which does exist
experimentally, finding excellent agreement a=5.836 (expt 5.865),
c=3.436 (expt 3.456), u=0.257 (expt 0.256), v=0.591 (expt 0.594).  This
gives us confidence that the ab initio calculations are reliable for
the system.  Results are given in Table.\ref{tab:crystal}.

\subsubsection{Monolayers and surfaces (table 1)}

To include fracture-relevant data, we
evaluated surface energies for pure Fe, Fe with a monolayer of P at
the surface, and Fe with a monolayer of P one layer below the surface
(there is evidence that P-rich grain boundary fracture occurs by
breaking of Fe-Fe bonds adjacent to the boundary, rather than Fe-P
bonds at the boundary.)  The results show that phosphorus does not 
segregate into a monolayer, either in bulk or on the surface, and that 
a free (001) surface with P on it has a higher energy than without (taking 
substitutional phosphorus as the reference state). Assuming replacement 
of iron by P, the segregation 
to grain boundaries must result from the different crystalline environment 
there rather than an intrinsic tendency of P to form layers (by extension, 
some grain boundaries will be more susceptible than others).  Moreover, 
the fracture embrittlement probably arises from a stress concentration effect 
rather than a simple Griffith-criterion energy balance.  We note however that  some 
experimental evidence\cite{bied} suggests that P sits in hollow sites on 
Fe(001) - investigating all such possible reconstructions by {\it ab initio} 
calculation would be impractical, hence the need for reliable potentials.

\subsubsection{Substitutional impurities (table 1)}

Phosphorus has a small range of solid solubility in Fe, and 
is believed to be located substitutionally.   Several 
configurations were
calculated: a single P atom in a 16(54) ($2^3$($3^3$) bcc) 
supercell and a 
pair of P atoms on neighbouring sites in similar cells.  In each 
case all ions were relaxed and we took the calculations to k-point 
convergence (729/216 k-points).  We compared the effects of fixing the 
lattice parameter at the pure Fe value, or relaxing it to minimise 
energy (see Table. \ref{tab:INT}).  The finite size effects going from 
16 to 54 atoms were only about 0.1eV, within the fitting errors.
In practice the Fe$_{15}$P-SSI value is used as the P reference state.

The addition of P impurities causes a striking reduction of 10-15\% in the 
stiffness of iron.  We calculate elastic constants by applying finite 
strains and measuring resultant stress.  Our calculations for pure iron 
give elastic constants of   C$_{11}$=225GPa, C$_{12}$=124GPa, C$_{44}$=101GPa, some 
10\% lower than experiment (and therefore not included in the fitting).  Calculations on the Fe$_{15}$P 
substitutional impurity supercell (6.25\% P) give  C$_{11}$=196GPa, C$_{12}$=109GPa, C$_{44}$=91GPa.

\subsubsection{Vacancies (table 2)}

With a 31-atom Fe unit cell, we find a vacancy formation energy of 
1.94eV compared to the fully converged value of 1.95eV\cite{domain}.  
With 30+1 atoms, the energy required to create a vacancy  
adjacent to a substitutional phosphorus is 1.64eV for the near neighbour 
site and 1.72eV for the second neighbour site.

For 15-atom supercells, the energies are 1.71eV for 
the near-neighbour site and
2.18eV for the second neighbour site, this latter 
suggests a P-vacancy repulsion, but the cell is so small 
that the calculation actually represents an
alternating chain P-vacancy-P-vacancy, so we neglect it.

We estimate the migration barrier by replacing two atoms in pure Fe with one 
located at their midpoint\cite{symm} and relaxing the remaining atoms while 
constraining the mirror plane symmetry.  As we shall see, in bcc this is not 
necessarily the barrier  for a ($\frac{1}{2}, \frac{1}{2}, \frac{1}{2}$) hop 
but it does represent a useful configuration 
to include in the fitting.
The energy for P to hop from the near neighbour site into the vacancy is 
much lower than for the direct second neighbour [001] hop: in the fitting 
we ensure that the latter barrier is high enough not to be surmountable in MD.

\subsubsection{Interstitials (table 2)}
In common with experiment\cite{Feintexp} and previous work\cite{domain,Fe} 
we find the [011]Fe-Fe dumbbell 
configuration to be the stable interstitial.  Geometrically, the migration 
mechanism for an [011] interstitial could go via the [111], tetrahedral, 
octahedral or [001] configuration - our calculations suggest the 
tetrahedral and [111] configurations are similar in energy.  Dynamically, 
it is also possible that a process of excitation to (111) followed by fast 
1D migration may be favoured\cite{Fe}.

Although we use smaller 
supercells without relaxation, the difference in energies from larger 
calculations can be approximated by elasticity\cite{thetford}, and the 
differences between various conformations (which is what we fit) 
are converged to within the accuracy with which we can fit them.   
When one or two phosphorus atoms 
form part of the interstitial the energy is lowered 
compared with an iron interstitial and substitutional phosphorus, and  
the stable configuration remains the [011] dumbbell.

\subsubsection{Liquid state calculations}

We use a self-consistent process to model liquids, starting with 
pair potentials\cite{fuji} and MD to create a ``typical''
atomic-level model of liquid Fe-P alloy (84 Fe atoms, 12 P).
 Then the total forces acting on each atom in this
model are obtained from static 
first principles calculations.   We then use dynamical refitting\cite{free} 
via force  matching\cite{adams} to produce a new trial 
potential and generate a new ``typical''
liquid configuration with classical MD. The process is then 
iterated to self-consistency. Liquid configurations provide data 
across the range of possible Fe-P separations, 
including small separations which are absent in equilibrium crystal 
data at T=0, and ensure there are no 
anomalous wiggles in the potential at separations for which there is no data. 

\subsection{Fitting}

In addition to the {\it ab initio} calculations above, other
properties of pure iron fitted to experiment are 
C$_{11}$=1.517eV/\AA$^3$, C$_{12}$=0.861eV/\AA$^3$,
C$_{44}$=0.761eV/\AA$^3$, and cohesive energy of 4.316eV.

We choose to parameterise the potential using a polynomial 
spline functional form.
This choice is arbitrary, but does not constrain the physical behaviour - we 
have shown\cite{mim} that a liquid simulation using a cubic spline fit can 
be used to fit a reciprocal power series potential, and the resulting $V(r)$ and $\phi(r)$ are indistinguishable.

Thus

\begin{equation}
 \phi(r_{ij}) = \sum_k A_k(R_k-r)^3 H(R_k-r)
\end{equation}
\begin{equation}
F(x)=-\sqrt{x}+a_2x^2+a_4x^4
\end{equation}

For the extreme short range repulsion, which is sampled only by the 
primary knock-on atom in a cascade simulation and not by any of our 
fitting data, we adopt the screened electrostatic form of Biersack and Ziegler\cite{bier}.

\begin{eqnarray}
V(r)=\sum_k  & a_k(r-r_k)^{n_k} H(r_k-r)H(r-r_2) \nonumber\\
            +& H(r_2-r)H(r-r_1)\exp{(B_0+B_1r+B_2r^2+B_3r^3)} \nonumber \\
+& H(r_1-r)\frac{Q_iQ_j}{r}\xi(r/r_s)
\end{eqnarray}

where $Q_i$ and $Q_j$ are the nuclear charges and 
where $r_s = 0.4683766/(Q_i^{2/3}+Q_j^{2/3})$.

\begin{equation}
\xi(x) = 0.1818e^{-3.2x} +0.5099e^{-0.9423x} +0.2802e^{-0.4029x} +0.02817e^{-0.2016x} 
\end{equation}

In these equations, $H(x)$ is the Heaviside function and $r_B$ 
the Bohr radius.  The $B_i$ coefficients are determined by continuity 
of the potential and its derivative, and so the parameters available
for fitting are the $A_k$, $a_k$ and C. 

We take $V_{FeFe}$, $\phi_{FeFe}$ and $F_{Fe}$ from our previous paper\cite{mimphilmag}. The $\phi$s 
are taken by scaling the pure iron values. The implicit assumption is that
in the rigid band picture, the tight binding energy goes
parabolically with number of valence d-electrons:  (10-N)N .  We
assume that in the dilute alloy the 3 phosphorus $p$-electrons end up 
in an unchanged d-band.  This suggests that the bond strength is e.g.
$\Phi_{FeP} / \Phi_{FeFe} = (3(10-3)/6(10-6))^2 = 0.765625$
the squaring coming from the assumption that the embedding function
 is a square root.
Obviously, this 
 means that the potential is invalid for high phosphorus concentration.

To fit the remaining parameters, we use a weighted least squares fit
to the lattice parameter and formation energy for several Fe$_3$P
compounds (D0$_3$, L1$_2$, stripe), the relative energies of defect
structures (including the $\langle 100 \rangle$, $\langle 110 \rangle$
and $\langle 111 \rangle$ mixed interstitial configurations), vacancy
-substitutional interaction and vacancy migration energies from the
first principles calculations.

In the weighting, prime importance was placed on those configurations
expected to be seen in simulation, the stable [110] interstitial, the
substitutional phosphorus and the migration barriers.  While other
configurations were less strongly weighted, we ensured that they
remained sufficiently high in energy that they would not subsequently
participate spuriously in dynamics.

\begin{figure}
\includegraphics[scale=1.0]{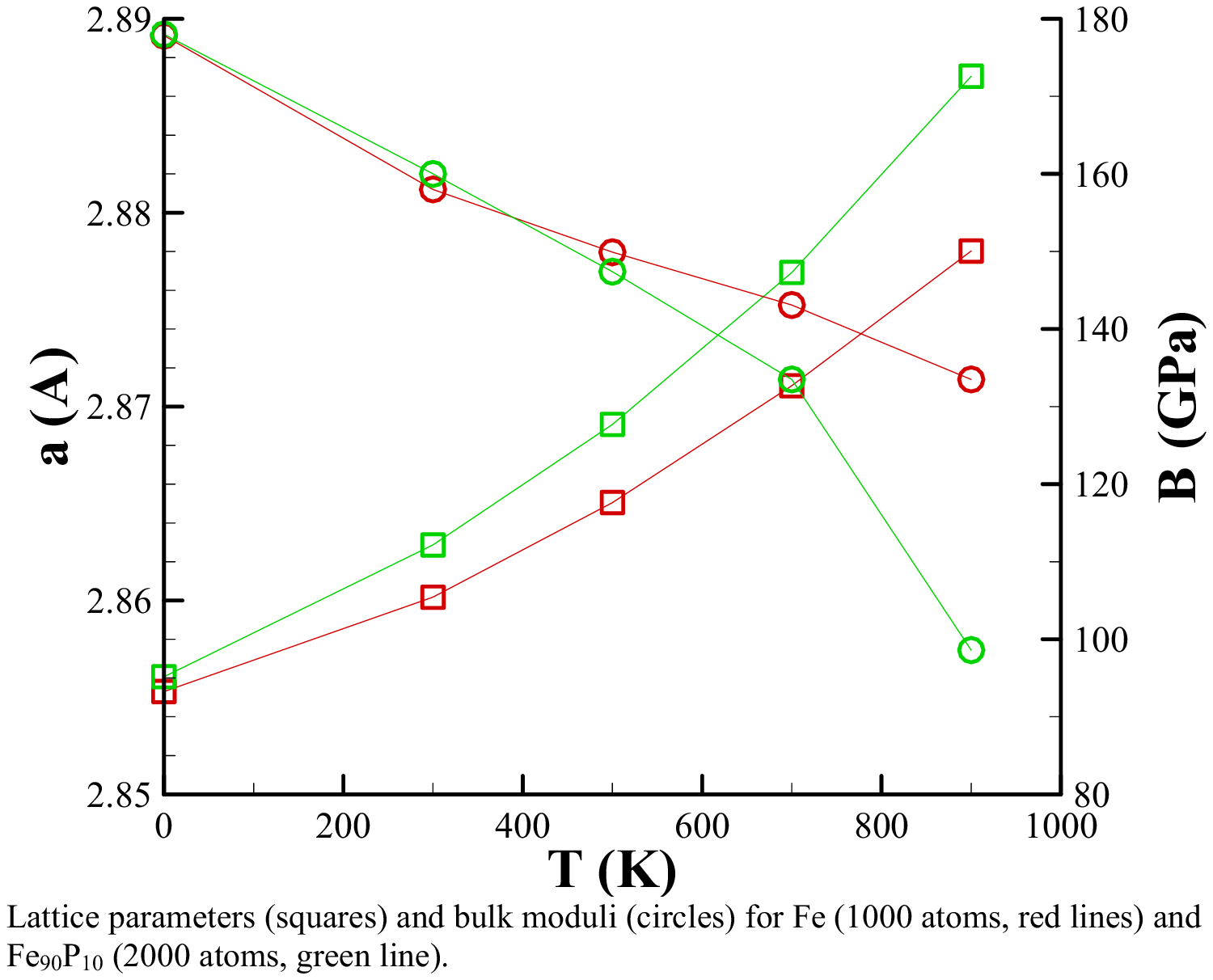}
\caption{Effect of 10\% P on Fe lattice parameter (squares) and 
bulk modulus(circles), evaluated by molecular dynamics on a 2000 
atom cell.  For comparison similar quantities for pure Fe (red lines) 
are shown. 
\label{fig:therm}}
\end{figure}

For the defects, as with the liquid, the fitting process is a 
self-consistent one - we fit to 
unrelaxed defect configurations to obtain a trial potential, 
relax the configuration using this potential, then 
refit the potential to the new configuration.  After several 
iterations, a self-consistent set of defect and liquid configurations, 
energies and potential parameters is obtained.  This process means 
that the interatomic distances found in the {\it ab initio} are 
not fitted.  Once a fit had been obtained, the functions $V(r)$ and $\phi(r)$ 
were tested {\it ab oculo} for overfitting and MD simulations for 
thermal expansion, elastic moduli (Fig.\ref{fig:therm}), 
vacancy and interstitial migrations and liquid state diffusion 
were done as a check for pathologies.  
The parameters for the final potential are given in table \ref{tab:para}.

\section{Phosphorus diffusion mechanisms}

\subsection{Interstitial Mechanism of Migration deduced from ab initio}

The models of P atom segregation proposed so far\cite{barbu,lidiard,barashev}
are developed for the case when the binding energy of a phosphorus-interstitial
complex is small.  Specifically that the mean free path
until thermally-activated dissociation is much shorter than
the mean distance between sinks of point defects.  This condition is
needed to justify the detailed balance approximation, required for the
concentration of P-interstitial complexes to be a function of
the local concentration of P atoms.  Our {\it ab initio}
calculations show that the binding energy of a
Fe-P mixed dumbbell is not small, and hence 
this approximation is not valid.  Indeed, the
large binding energy and small migration energy of an Fe-P 
interstitial complex implies that during the lifetime of this
complex the probability that it will dissociate thermally is
small.  
For a typical value of phosphorus concentration in RPV
steels, $\sim 10^{-3}$, the mean free path of an irradiation-produced Fe-Fe dumbbell before encountering a P atom is 
several nanometres. Hence, the
majority of interstitials arriving at a GB (or any other sink of
point defects) will contain phosphorus atoms.  As a consequence of 
the phosphorus-interstitial complex should be treated as a single 
migrating entity in methods such as kinetic Monte Carlo.

\subsection{Vacancy Mechanism of Migration} 

Our results also
show significant vacancy-P atom interaction energy: the binding energy
of this complex to be $\sim 0.3$eV.  This was not expected previously and
in the model by Lidiard\cite{lidiard} this interaction was totally neglected. 
Even more importantly the interaction is long-ranged, at least up to the
second neighbour atoms, which invalidates the simple 
 models\cite{barbu} for diffusion coefficients in Fe-P
alloy. Their conclusion, that in bcc alloys solute
atoms always drift up the vacancy concentration gradient\cite{barbu},
must be re-examined allowing for longer ranged interactions and
complex formation. The long-range vacancy-phosphorus interaction in 
bcc iron makes it possible
for a vacancy to move around the P atom, while remaining bound as a complex.  
As a consequence the situation may become similar to fcc alloys, where the
diffusion coefficient of a solute atom can be positive (indicating
drag of solute atoms) or negative (simple exchange) depending on the
relative frequencies of vacancy jumps between two neighbour sites of
the solute atom and away from the solute atom.  Hence, it is 
possible that vacancies would drag P atoms to sinks of point defects.  

The small energy for the vacancy P atom
exchange jump (relative to that of vacancy iron atom) implies that the
diffusion coefficient of phosphorus atoms via a vacancy mechanism 
would be independent of this energy.  The potential shows that
the barrier for first-to-second neighbour jumps is similar to barriers 
in pure Fe.  Finally, the conclusion\cite{lidiard} that the
migration of phosphorus atoms to grain boundaries is predominantly via
the interstitials has to be verified in the view of strong long
ranged interaction of vacancy P atom complexes, and smaller production
rates of single interstitials than single vacancies in high-energy
displacement cascades in $\alpha$-iron due to higher intra-cascade
clustering of interstitial atoms.

\subsection{Calculated Mechanism of Migration using the Potential} 

\begin{figure}
\includegraphics[scale=1.0]{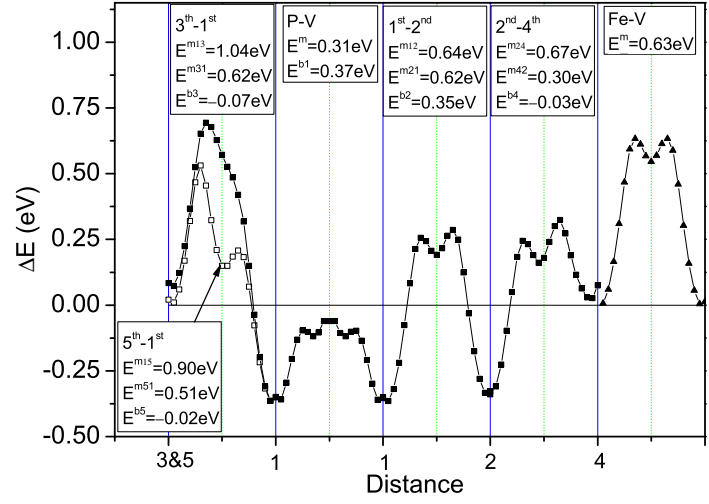}
\caption{Constrained static relaxations of P-vacancy complex with P
moved over various barriers.  $E^{mxy}$ gives the height of the
barrier for the P to move from $x^{th}$ neighbour to the $y^{th}$,
$E^{bx}$ gives the binding energy of P at the $x^{th}$ neighbour
site. The final panel shows the barrier for vacancy movement in pure
iron.  The barrier against motion of the 
vacancy-phosphorus complex is similar to the binding energy, 
and much lower than the barrier in pure Fe. Consequently, the P 
traps the vacancy, reducing overall vacancy diffusion.\label{fig:vhop}}
\end{figure}

We have performed simulations of diffusion for point defects in FeP
using the potential defined in table.\ref{tab:para}.  We find that
vacancies diffuse freely in iron, but are attracted to and form
complexes with the P atoms.  Although the phosphorus atom can move
into the vacant site, in isolation this would simply produce a
thermally activated oscillation, and no net diffusion - the vacancy
has to hop out to second neighbour sites and back again for
comigration to occur.  Using static relaxation, we calculated the various 
barriers to vacancy jumps in the vicinity of a P atom (Fig.\ref{fig:vhop}).  
A near-neighbour 
hop in bcc crosses two intermediate (111) planes, so the barriers tend to 
be bimodal.  Although the P-v migration barrier is only half that of the 
Fe-v barrier, to obtain long range P migration the vacancy has to hop 
around the P via second neighbours.  Energetics (both ab initio and 
with the potential) show the P to be bound to the vacancy in both 
first and second neighbour sites, however from the second neighbour site 
the barriers are similar for hopping back or further away from the P.
Thus  P acts as a strong vacancy trap, and comigration is likely. 

\begin{figure}
\includegraphics[scale=0.8]{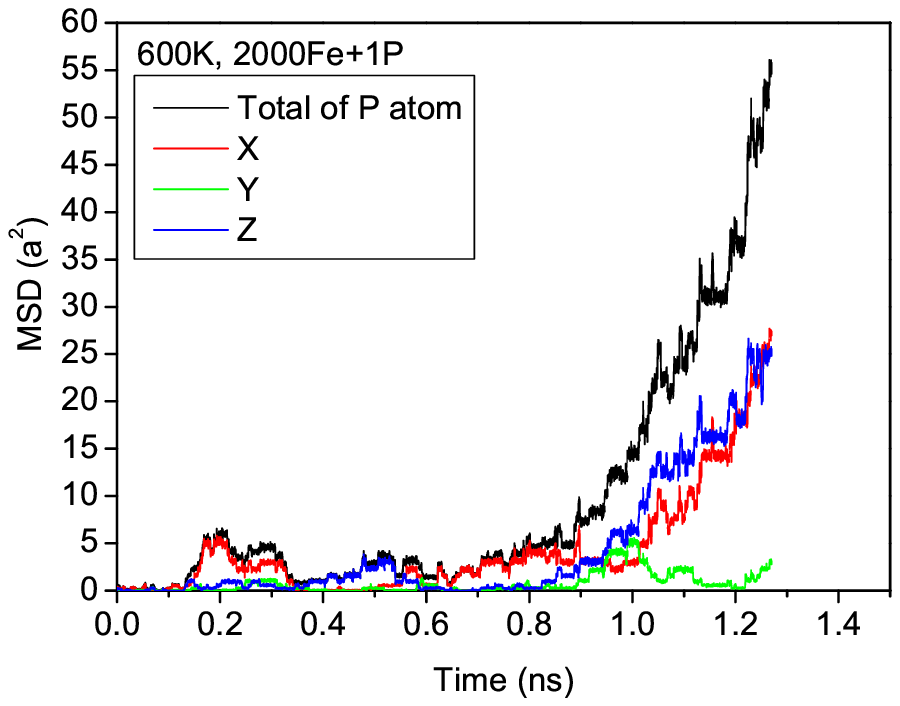}
\includegraphics[scale=0.8]{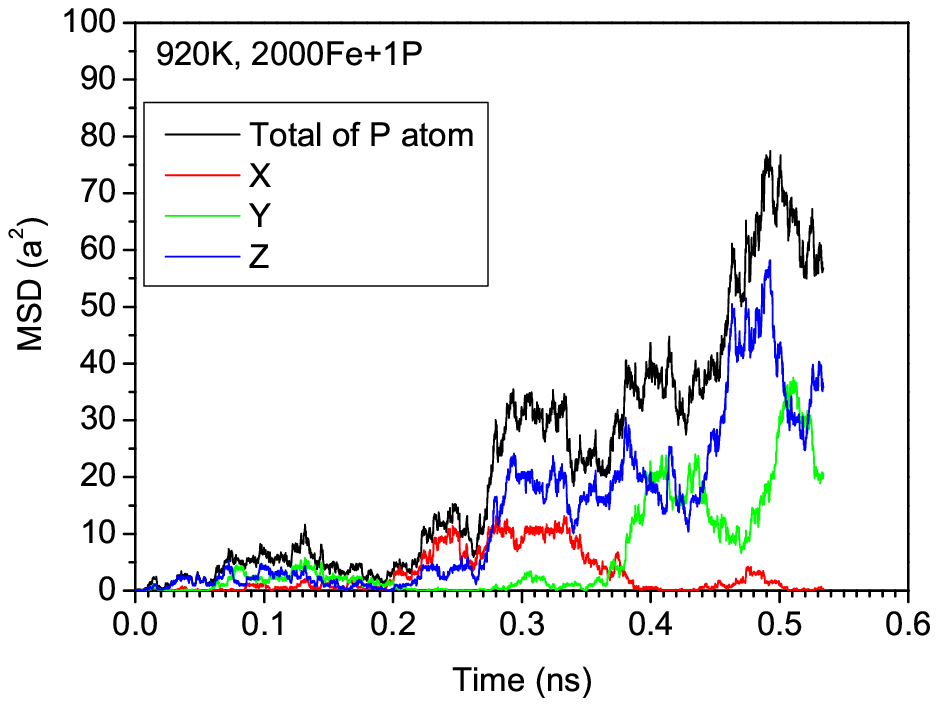}
\includegraphics[scale=0.8]{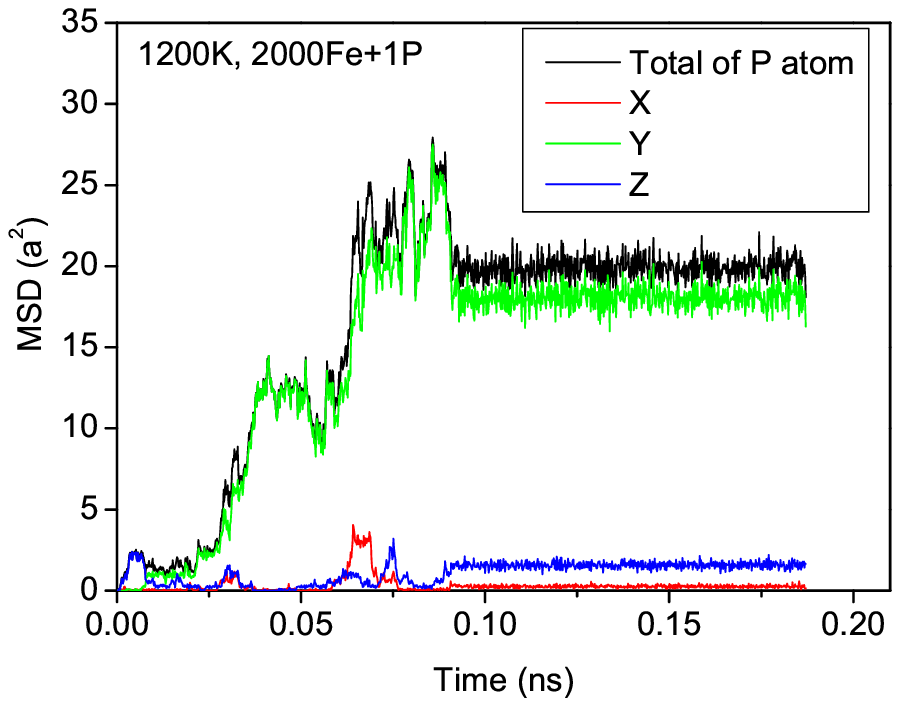}
\includegraphics[scale=0.8]{md1200fe.eps}
\caption{Mean Squared Displacement (MSD) against time for migration in a
unit cell containing 2000 Fe and 1P atom, initially located in a mixed
dumbbell.  Various temperatures were examined: at 600K and 920K the P
remains attached to the interstitial throughout the period of the run.
For 1200K, we observed rapid 1D diffusion of the phosphorus atom via 
mixed interstitial (Fe does not diffuse in this phase),
followed by a dissociation event and standard interstitial diffusion in 
pure Fe.  The simulation box is large enough that the interstitial is 
not recaptured, and subsequent diffusion is for the iron atoms. 
\label{bara:mig}}
\end{figure}

We performed some preliminary MD simulations using the potential with 2000 
atoms to check the mechanism and ensure against pathological behaviour.  
For 1200K, with a single P interstitial, a series of 1D migration steps were 
observed for about 80ps, after which the interstitial dissociated from the 
P atom and standard 3D migration in pure Fe continued.  At 920K and 600K, 
however, the mixed interstitial did not separate and 
migration was more three dimensional (Fig.\ref{bara:mig}).  Interstitials 
diffuse freely in pure iron, and
are strongly attracted to substitutional P.  The P-Fe mixed dumbbell
is also highly mobile, and the P diffuses rapidly. Thus P-diffusion
occurs primarily via interstitials in radiation damage conditions
where Fe-Fe interstitials are commonplace. Under thermal conditions
however, interstitial mediated segregation to boundaries will not
occur since there is no source for interstitials in the bulk, and
mixed interstitials formed at sinks cannot transport further P-atoms
to the sink.

The stability of small complexes suggests, however, that still larger
complexes may be important. For example mixed interstitials 
may be pinned by other P atoms, forming sessile PP interstitials. 
Future molecular dynamics can incorporate
features neglected by this model, such as clustering, recombination,
segregation and the specific defect distributions associated with
thermal, electron or neutron radiation.

\section{Conclusions}

We have developed an interatomic potential for the $\alpha$-Fe-P
system in the limit of low-P concentration with particular importance
attached to configurations observed under radiation damage.  This
potential gives a good description of point defect properties and
conformations, in particular vacancy and interstitial complexes with a
P atom and hence can be used in larger scale molecular dynamics or
Monte Carlo simulations to study diffusion properties and segregation
of P during both ageing and irradiation conditions.  Pure phosphorus
cannot be described by this type of potential.

The potential predicts that phosphorus 
is strongly bound to both vacancy and 
interstitial defects, in accordance with 
{\it ab initio} results.  This fact has not been expected previously, and conclusions of theoretical studies that ignore strong P-defect binding  should be reexamined.

We have presented prelimnary molecular dynamics studies of P diffusion
in iron which show the potential to be free from pathologies, but a
quantative study of these processes (which should include larger
complexes) in beyond the scope of this paper.  It appears
that in the temperature region of interest for RPV steels P would
migrate rapidly via both interstitial and vacancy complex mechanisms, 
although we have not studied large complexes which may further 
complicate the process.  

In summary, the detailed understanding of radiation damage in steels
remains an area of active interest more than thirty years after the
pioneering work of Michael Norgett helped pose the questions.

\subsection*{Acknowledgements}

AVB acknowledges financial support from the European Commission via contract
FIKS-CT-2000-00080 ("PISA"),  GJA from framework 6 project PERFECT.

\newpage
\begin{table}[h]


\begin{tabular}{|l|l|l|l|l|l|l|}

\hline
Formula & Structure & Energy (eV) & Mag.Mom & Volume A$^3$ & Pressure \\
\hline
Fe$_2$  &  bcc  &  16.617 & 4.62 & 23.24 & 0 \\
Fe$_3$P & L1$_2$ & 30.755 & 7.85 & 47.59 & 0 \\
Fe$_3$P & DO$_3$ & 30.971 & 5.46 & 43.55 &0\\
Fe$_3$P & stripe & 30.786 & 5.52 & 44.69 &0\\
Fe$_6$P$_3$ & Fe$_2$P & 69.718 & 9.13 & 101.34 &0\\
\hline
Fe$_{15}$P & SSI & 130.867 & 35.30 & 185.9   & 1.34  \\
Fe$_{53}$P & SSI & 446.714 & 125.26 & 630.4   & 0  \\
Fe$_{14}$P$_2$ & DSI & 128.407 & 32.71   & 185.9  &  6.67 \\
\hline
\hline
Formula & Strain & Energy (eV) & Mag.Mom & Volume A$^3$ & Stresses\\
\hline
Fe$_7$P &  $\epsilon_{zz} =$ -0.025& 
63.997& 14.6  & 88.24 & 65, 82 \\
Fe$_7$P &  $\epsilon_{zz} =$ 0.0 & 
64.066& 14.8 & 90.51   &  25,  10\\
Fe$_7$P &  $\epsilon_{zz} =$ 0.025 & 
64.042 & 15.0 &92.77   & -13,   -51\\
Fe$_7$P &  $\epsilon_{zz} =$ 0.050  & 
63.941 &15.2  &95.03 &  -45,   -98\\
Fe$_7$P &  $\epsilon_{zz} =$ 0.075 & 
63.789 &15.6 & 97.30 &  -70  -130 \\
Fe$_7$P &  $\epsilon_{zz} =$ 0.5  & 
61.073 & 10.9 & 135.76 & -26, 0  \\
Fe$_6$PFe &  $\epsilon_{zz} =$ 0.5  & 
60.988 & 6.9 & 135.76 & -24, 0 \\
Fe$_8$ &  $\epsilon_{zz} =$ 0.5  &  64.141  & 21.0 & 135.76 & -6, 0 \\

\hline
\end{tabular}
\caption{Energies of relaxed iron-rich crystal structures (``stripe'' 
is a layer bcc structure with Fe-Fe-Fe-P (001) layers), single 
substitutional impurity phosphorus atoms (SSI), near-neighbour double 
substitutional impurity (DSI), and an (001) monolayer of P in 7 layers of Fe, 
(labelled Fe$_7$P) both under strain and after fracture at the P (Fe$_7$P),
one layer above the P (Fe$_6$PFe) and in pure iron (Fe$_8$).  Stresses are
parallel and perpendicular to the long (z) axis respectively.
 The quoted
energies are relative to a spin-zero GGA representation of an iron
atom. Absolute values are not used in the fitting, only differences
between them for which this arbitrary zero cancels out. 
\label{tab:crystal}
}

\end{table}

\begin{table}
\begin{tabular}{|l|l|l|l|l|l|l|l|}

\hline
Formula & Structure & Energy (eV) & Mag.Mom & Volume(A$^3$) & Pressure (kB) &adjusted energy (eV)\cite{thetford} \\
\hline

 Fe$_{15}$P$_2$ & 111 &  132.470/4.636 &   33.8 & 185.9  & 218 &2.91 \\
 Fe$_{15}$P$_2$ & 001 &  132.464/4.642  &   33.3 & 185.9 &  246 &2.45\\
\bf Fe$_{15}$P$_2$& 011 & 133.419/3.687 &  34.1 &  185.9 &   223 &1.88  \\\rm
Fe$_{16}$P & TET & 134.828/4.347 & 32.8 & 185.9 &    203  &2.85 \\
Fe$_{16}$P & 001 & 134.428/4.747 & 35.8  & 185.9 &   252  &2.45   \\\bf
Fe$_{16}$P & 011 & 135.353/3.822 & 32.7 & 185.9  &   198 &2.39 \\\rm
Fe$_{16}$P & 111 & 134.820/4.355 & 35.7 & 185.9  &   224 &2.54 \\
Fe$_{16}$P & OCT & 134.428/4.747 & 35.8  & 185.9 &    252 &2.45 \\
Fe$_{17}$  & TET & 135.723/5.521 & 31.4  & 185.9 & 184 &4.29   \\\bf
Fe$_{17}$  & 011 & 136.419/4.825 & 34.2  & 185.9 &  201 &3.36 \\\rm
Fe$_{17}$  & 001 & 134.779/6.466 &  34.6 & 185.9 &  224  &4.65 \\
Fe$_{17}$  & 111 & 135.746/5.498 &  34.9  & 185.9  &  201 &4.03 \\
Fe$_{17}$  & OCT & 134.530/6.715 & 35.0  & 185.9 & 228    &4.83 \\
\hline
\end{tabular}
\caption{Calculated results for interstitial formation energies,
relative to GGA free atoms / pure iron and SSI phosphorus.  Fe$_{17}$ denotes iron interstitials, Fe$_{16}$P
denotes mixed dumbbells or P interstitials, Fe$_{15}$P$_2$ denotes P-P
dumbbells. TET and OCT denote tetrahedral and octahedral sites respectively, 
other structures are dumbbell configurations.  
In each case volume is constrained to the 16-atom pure
iron cell, a 9x9x9 k-point grid was used and all atomic positions
relaxed.  The mixed dumbells have low symmetry and may relax to a 
higher symmetry state: [001] mixed dumbell relaxes to the octahedral 
position and the [111] mixed dumbell goes to the crowdion position.
These calculations are for small supercells incorporating both
interstitial formation energy and strain interactions - they should
not be regarded as energies of isolated defects (they are probably 
upper bounds).  Cohesive energies
are quoted firstly relative to non-magnetic free atoms with GGA
secondly relative to solid Fe and substitutional Fe$_{15}$P.  
Elastic correction for finite size\protect{\cite{thetford}}
of $P^2V/2B$  suggests
that full relaxation in a large supercell would lower the formation
energies systematically by about 1.5eV
\protect{\label{tab:INT}} }
\end{table}

\begin{table}
\begin{tabular}{|l|l|l|l|l|l|l|}

\hline
Formula & Structure & Energy (eV) & Mag.Mom & Volume (A$^3$) & Pressure(kB)  \\
\hline

Fe$_{15}$ & vacancy &    122.606/2.021  &  36.4 &   183.66   &   0  \\
Fe$_{15}$ & bar vacancy &    122.008/2.619  & 36.4   &   184.56   &   0  \\
Fe$_{14}$ & divac-1st &  112.563/3.757 &  34.9 & 180.20   &  0  \\
Fe$_{14}$ & divac-2nd &  112.838/3.481 &  35.5 & 179.10  &   0  \\
Fe$_{14}$P & 1st  &   120.852/1.706  &  33.2 &  181.78  &   0  \\
Fe$_{14}$P & 2nd  &   120.379/2.180  &  34.6 &  184.64  &    0  \\
Fe$_{14}$P & bar-111  &   120.504/2.055  &   33.7 & 184.14   & 0 \\
Fe$_{30}$P & 1st  &   253.873/1.622  &  71.1  &   371.87   &    -15  \\
Fe$_{30}$P & 2nd  &   253.775/1.719  &   72.3 &   371.87   &     -7    \\
Fe$_{30}$P & bar-100  & 252.006/3.489   &   70.8 & 371.87 &   -19  \\
Fe$_{30}$P & bar-111  & 253.525/1.970   &   71.1 & 371.87 &   -8   \\
\hline
\end{tabular}
\caption{Calculated results used to fit 
vacancy formation energies, relative to GGA free atoms/pure iron 
and SSI phosphorus.  Formulae give number of atoms in the
supercell, ``divac'' represents two vacancies at nearest and second 
neighbour sites 1st and 2nd represent the site of the P 
relative to the vacancy, ``bar'' represent the energy with the P 
midway along its migration path.
In 15 atom cases unit cell and atomic positions are fully relaxed, 
and a 9x9x9 k-point grid used.  These vacancy-rich configurations 
incorporate both vacancy formation energy and strain interactions 
- they should not be regarded as energies of isolated defects.  
In particular, for the second neighbour cases we have a chain of second 
neighbour vacancies.
Comparison with results for larger supercells\cite{domain} suggests that 
full relaxation in a large supercell would lower the energies slightly, but maintain the differences between them.
\label{tab:v}}
\end{table}

\begin{table}
\begin{tabular}{|l|l|l|}

\hline
 Structure & Energy (eV) & Dilatation(\%)  \\
\hline
pure iron & 4.013 & 0 \\
Fe vacancy & 1.71 & -22.3\\
Divac-1st & 3.285 & -35.3\\
Divac-2nd & 3.18 &  -48.9\\
Bar-Vacancy-Fe & 2.34 & - \\
Int 110 & 3.59 & +124.7\\
Int OCT & 4.22 & +102.1\\
Int 100 & OCT &\\
Int 111 & 110 &\\
\hline
P SSI & 0.0 & -34.3\\
P-vac (1st) & 1.34 & -57.1\\
P-vac (2nd) & 1.37 & -58.1\\
Bar P-vac & 1.65 & - \\
P-110 & 2.57 & +115.7\\
P-111 & 3.30& +102.7\\
P-100 & Int 110 & \\
P-TET & 2.80 & +151.5\\
P-OCT & 3.47 & +161.5\\
Bar P-SSI-110 & 0.27 & - \\
\hline
\end{tabular}
\caption{Energies and dilatations of defect configurations calculated
using the interatomic potential, with 2000 atom constant (zero).  
pressure static relaxation. Notation is as for previous tables.
Energy refers to a reference state with an equivalent number of 
iron atoms in pure iron and phosphorus atoms as SSIs in iron.  For 
unstable interstitial configurations we indicate the local minimum.
Dilatation is defined relative to pure iron and calculated using 
constant volume and elastic constants\cite{schober}
\label{tab:pot}}
\end{table}

\newpage
\begin{table}
\begin{tabular}{|l|l|l|}
\hline\small
Potential & Value (eV) & Cutoffs ($\AA$) \\
\hline\hline
$V_{FeFe}(r)$ &$9734.2365892908\xi(r,26,26)/r$ & 0.0-1.0 \\&$
+\exp(7.4122709384068-0.64180690713367r-2.6043547961722r^2+0.6262539393123r^3)
$&1.0-2.05\\&$-27.444805994228(2.2-r)^3
$&2.05-2.2\\&$+15.738054058489(2.3-r)^3
$&2.05-2.3\\&$+2.2077118733936(2.4-r)^3
$&2.05-2.4\\&$-2.4989799053251(2.5-r)^3
$&2.05-2.5\\&$+4.2099676494795(2.6-r)^3
$&2.05-2.6\\&$-0.77361294129713(2.7-r)^3
$&2.05-2.7\\&$+0.80656414937789(2.8-r)^3
$&2.05-2.8\\&$-2.3194358924605(3.0-r)^3
$&2.05-3.0\\&$+2.6577406128280(3.3-r)^3
$&2.05-3.3\\&$-1.0260416933564(3.7-r)^3
$&2.05.3.7\\&$+0.35018615891957(4.2-r)^3
$&2.05-4.2\\&$-0.058531821042271(4.7-r)^3
$&2.05-4.7\\&$-0.0030458824556234(5.3-r)^3$
& 2.05-5.3\\
\hline
$V_{FeP}(r)$&$
(5615.9057245908/r)\xi(r,26,15)
$&0.0-1.0\\&$
+exp(10.76185442488-10.004045788895r+4.9854254472397r^2-1.2599788569372r^3)
$&1.0-2.0\\&$
-3.3136605743629(5.3-r)^4$&2.0-5.3\\&$
+12.62523819360(5.3-r)^5$&2.0-5.3\\&$
-20.361693308072(5.3-r)^6$&2.0-5.3\\&$
+17.629292543942(5.3-r)^7$&2.0-5.3\\&$
-8.8120728047659(5.3-r)^8$&2.0-5.3\\&$
+2.5494288609989(5.3-r)^9$&2.0-5.3\\&$
-0.39698390783403(5.3-r)^{10}$&2.0-5.3\\&$
+0.025779015833433(5.3-r)^{11}$&2.0-5.3
\\ \hline
$V_{PP}(r)$& $(3239.9456103409/r)\xi(r,15,15)$&0.0-0.9\\&$
+exp(9.9382842499617-8.5637164272526r+3.451962728599r^2-0.61453831350215r^3)
$&0.9-2.5\\&$
-0.078293794709143(5.3-r)^4$&2.5-5.3\\&$+0.037557214911646(5.3-r)^5
$
& 2.5-5.3
\\
\hline
\hline
$\phi_{FeFe}(r)$&
$ 11.686859407970(2.4-r)^3
$&0.0-2.4\\&$-0.01471074009883(3.2-r)^3
$&0.0-3.2\\&$+0.47193527075943(4.2-r)^3$&
0.0-4.2\\
\hline
$\phi_{FeP}(r)$&$\phi_{FeFe}(21/24)^2$& 0.0-4.2\\
\hline
$\phi_{PP}(r)$&$\phi_{FeFe}(21/24)^4$ &0.0-4.2\\
\hline
\hline
$F_{Fe}(\rho)$& 
$-\sqrt{\rho}-6.7314115586063 \times 10^{-4}\rho^2+7.6514905604792 \times 10^{-8}\rho^4$
&\\
\hline
$F_P(\rho)$& $-\sqrt{\rho}+0.0011950274540243\rho^2$&\\
\hline
\end{tabular}
\caption{Analytic form of the potentials.  $\xi$
is the short-range screening function\cite{bier} described in the text.
As further work is done, we intend to continue the 
self-consistent fitting process to improve the potential.
Users are encouraged to contact the authors with results and an evolving 
``best'' set of parameters will be maintained online at http://homepages.ed.ac.uk/graeme
\label{tab:para}}
\end{table}


\begin{thebibliography}{99}
\bibitem{symm} ($\frac{1}{4},\frac{1}{4},\frac{1}{4}$) 
is the symmetric midpoint, however if no relaxation occurs 
the migrating atom makes its closest approach to other neighbours at 
($\frac{1}{6},\frac{1}{6},\frac{1}{6}$) and 
($\frac{1}{3},\frac{1}{3},\frac{1}{3}$).
\end{thebibliography}
\end{document}